# Superconductivity in $Fe_{1+d}Te_{0.9}Se_{0.1}$ induced by deintercalation of excess Fe using alcoholic beverage treatment

K. Deguchi[1,2], S. Demura[1,2], H. Hara[1,2], T. Watanabe[1,2], S. J. Denholme[1], M. Fujioka[1], H. Okazaki[1], T. Ozaki[1], H. Takeya[1], T. Yamaguchi[1] and Y. Takano[1,2]

[1] *National Institute for Materials Science, 1-2-1, Sengen, Tsukuba, 305-0047, Japan*

[2] *University of Tsukuba, 1-1-1 Tennodai, Tsukuba, 305-8571, Japan*

E-mail: DEGUCHI.Keita@nims.go.jp

Abstract

Superconductivity in polycrystalline $Fe_{1+d}Te_{0.9}Se_{0.1}$ was enhanced by heating in alcoholic beverages. We found that part of the excess Fe was deintercalated from the interlayer sites by the treatment and the shielding volume fractions were increased with the reduction of excess Fe content in sample. This behavior is similar to the case of $Fe_{1+d}Te_{1-x}S_x$. Thus the technique of alcoholic beverage treatment can be applied generally to the Fe-chalcogenide superconductors.

## 1. Introduction

Since the discovery of superconductivity in LaFeAsO$_{1-x}$F$_x$, research into Fe-based superconductivity has been actively performed and several types of Fe-based superconductors have been discovered [1-4]. All the Fe-based superconductors share a common layered structure based on a planar layer of Fe square lattice. In Fe-pnictides superconductors, blocking layers with alkali, alkali-earth or rare-earth and oxygen/fluorine are alternatively stacked with Fe-As conduction layers. Contrary to the Fe-pnictides, Fe-chalcogenides superconductors are composed of only superconducting Fe-Se and -Te layers. Parent compounds of Fe-chalcogenides are FeSe and FeTe. The superconducting transition temperature of the FeSe ($T_c \sim$ 8K) can be raised to 37 K by applying pressure and to 14 K by partial Te substitution for Se [5-10]. In contrast, FeTe is not superconducting. Instead it exhibits a simultaneous structural and antiferromagnetic phase transition at 70 K.

Previous studies have demonstrated that the spin structure of FeTe is different from that of the Fe-pnictide parent compounds [11]. For FeSe and Fe-pnictide superconductors, the antiferromagnetic wave vector $Q_s$ = (0.5, 0.5), which correlates with superconductivity, was confirmed [12-16]. On the other hand, FeTe shows magnetic wave vector $Q_d$ = (0.5, 0) [12, 13]. Several reports indicated that the wave vector $Q_d$ is not favorable for superconductivity [17-19]. The emergence of a Fermi surface nesting associated with $Q_d$ could be induced by excess Fe at the interlayer sites. The excess Fe donates charge to the FeTe layers acting as an electron dopant and is strongly magnetic [17, 19]. FeTe can only be synthesized with large amounts of excess Fe. Thus the physical properties of $Fe_{1+d}Te$ depend on the excess Fe content $d$. The substitution of Te by Se reduces the content of excess Fe and suppresses the low temperature structural/magnetic phase transition [5, 20]. However $Fe_{1+d}Te_{1-x}Se_x$ samples with lower $x$ are still greatly affected by excess Fe due to the higher content of $d$. For $Fe_{1+d}Te_{0.9}Se_{0.1}$, the long-range magnetic ordering was not suppressed and only a weak superconducting signal was observed from magnetic susceptibility measurements.

So far vigorous attempts have been carried out to control the effect of excess Fe in $Fe_{1+d}Te_{1-x}Se_x$. Y. Kawasaki *et al*. reported that oxygen annealing suppressed the magnetic ordering and induced bulk superconductivity [21]. Oxygen ions intercalated between the superconducting layers suppressed the magnetic wave vector $Q_d$ due to the compensation of the electron given by the excess Fe. Furthermore, in the case of $Fe_{1+d}Te_{1-x}S_x$, we found that alcoholic beverage heating suppresses the effect of excess Fe by the reduction in the content of $d$ and induces superconductivity [22-24]. $Fe_{1+d}Te_{0.8}S_{0.2}$ heated in red wine at 70 °C for 24 h showed zero resistivity at 7.8 K and a shielding volume fraction of 62.4 %, whereas the as-grown sample did not. A technology of metabolomics analysis revealed that the weak acid in alcoholic beverages has the ability to deintercalate part of the excess Fe from the interlayer sites of sample and as a result superconductivity is achieved. Alcoholic beverage heating is an easy way to achieve the reduction of excess Fe content in the sample. However, this unique technique was not confirmed for $Fe_{1+d}Te_{1-x}Se_x$. Here we show the enhancement of superconductivity in polycrystalline $Fe_{1+d}Te_{0.9}Se_{0.1}$ by immersing the sample in alcoholic beverages.

**2. Experiment**

Polycrystalline samples of $Fe_{1+d}Te_{1-x}Se_x$ with $x = 0.1$ were prepared by a solid-state reaction using powder Fe (99.998 %) and grains of Te (99.9999 %) and Se (99.999 %). The starting materials with a nominal composition of $FeTe_{0.9}Se_{0.1}$ and a weight of about 1.0 gram were put into a quartz tube. The quartz tube was then evacuated by a rotary pump and sealed. After being heated at 650 °C for 10 hours, the obtained mixture was ground, weighed and pelletized into separate pellets with an approximate weight of about 0.100 grams each. The pellets were sealed into an evacuated quartz tube and heated at 650 °C for 10 hours. After furnace cooling, the obtained pellets were put into a glass bottle filled with 10 ml water and the alcoholic beverages: red wine (Bon Marche, Mercian Corporation), white wine (Bon Marche, Mercian Corporation), whisky (The Yamazaki Single Malt Whisky, Suntory Holdings Limited), sake (Hitorimusume, Yamanaka shuzo Co., Ltd.), beer (Asahi Super Dry, Asahi Breweries, Ltd.), or shochu (The Season of Fruit Liqueur, Takara Shuzo Co., Ltd.). pH values of the red wine, white wine, whisky, sake, beer, shochu and water were 3.38, 2.92, 3.88, 4.99, 4.07, 7.70 and 7.00 respectively. The samples in various liquids were heated at 70 °C for 24 hours. Powder X-ray diffraction patterns were measured using the $2\theta/\theta$ method with Cu-$K\alpha$ radiation by Mini Flex II (Rigaku). The temperature dependence of magnetization was measured using a Magnetic Property Measurement System (Quantum Design) magnetometer down to 2 K after both zero-field-cooling (ZFC) and field-cooling (FC) with an applied field of 10 Oe. We defined $T_c^{mag}$ as an onset temperature of the drop. The shielding volume fraction was estimated from the value of the magnetic susceptibility at 2 K after ZFC. Fe concentrations of the solutions were analyzed using inductively-coupled plasma (ICP) spectroscopy (Thermo Scientific).

### 3. Results and discussion

Temperature dependence of magnetic susceptibility for various samples is shown in

Figure 1. All the samples heated in solution show superconductivity, whereas the diamagnetic signal corresponding to superconductivity is not observed in the as-grown sample. The alcoholic beverage samples exhibit almost the same $T_c^{mag}$ of 10.3 K. In contrast, $T_c^{mag}$ of the water sample is about 9.0 K. We estimated the shielding volume fractions of the samples heated in the red wine, white wine, whisky, sake, beer, shochu and water to be 64.3, 56.8, 48.9, 46.0, 43.7, 32.2 and 23.7 %, respectively. The obtained shielding volume fractions are summarized in Figure 2 as a function of pH. The highest volume fraction was observed for the red wine sample and the smallest was obtained with the shochu sample. The alcoholic beverage samples have a higher value of $T_c^{mag}$ and shielding volume fraction compared with the water sample under the same heating conditions. This indicates that the alcoholic beverages can enhance the superconductivity in not only $Fe_{1+d}Te_{1-x}S_x$ but also $Fe_{1+d}Te_{1-x}Se_x$.

Our previous work reported in ref 23 suggests that heating $Fe_{1+d}Te_{1-x}S_x$ in acid solutions deintercalated part of the excess Fe from the interlayer sites of the sample, and then induced superconductivity. To investigate the effect of alcoholic beverages for $Fe_{1+d}Te_{1-x}Se_x$, ICP measurements were performed to analyze the Fe concentrations in the solutions. The average concentration of Fe in the treated red wine, white wine, whisky, sake, beer, shochu, and water are 50.5, 46.2, 11.7, 19.1, 17.8, 0.2 and 1.3 ppm.

Before treatment, only the red and white wine contained a small amount of Fe ions with concentrations of 2.3 and 2.8 ppm respectively. Fe ions were not detected in the other liquids. This result indicates that the Fe ion in the solutions was removed from the $Fe_{1+d}Te_{0.9}Se_{0.1}$ sample by heating in alcoholic beverages.

Figure 3 shows X-ray diffraction patterns for the as-grown $Fe_{1+d}Te_{0.9}Se_{0.1}$ sample and the samples heated in liquid. We found that heating the samples in these solutions does not degrade the crystal structure. The variations of the calculated lattice constants $a$ and $c$ between the samples are less than 0.2 %. There is no significant difference in the X-ray diffraction patterns between the as-grown sample and the treated samples at the sensitivity range of lab-level X-ray powder diffraction.

As mentioned above, the crystal structure of the samples was not decomposed by treatment. In fact, ICP analysis indicates that all treated solutions did not contain Te and Se ions. Therefore, as with the case of $Fe_{1+d}Te_{1-x}S_x$, the excess Fe was deintercalated from the interlayer sites of $Fe_{1+d}Te_{1-x}Se_x$. We summarize the relationship between the shielding volume fraction and content of the dissolved excess Fe from $Fe_{1+d}Te_{0.9}Se_{0.1}$ samples in Figure 4. It is obvious that the shielding volume fractions were increased with the reduction of excess Fe in the sample. According to previous reports, excess Fe in the $Fe_{1+d}Te_{1-x}Se_x$ with a lower $x$ is about 2 ~ 8 mol% [25, 26]. In contrast, about 1.6 mol% of excess Fe was deintercalated using red wine at 70 °C for 24 h. Furthermore, the red wine sample shows a $T_c$ of 10.4 K, this is slightly low when compared with the 12.8 K value obtained from oxygen annealed $Fe_{1+d}Te_{0.9}Se_{0.1}$ which suppressed the effect of excess Fe. We clearly demonstrate that the excess Fe which exists unavoidably in the as-grown $Fe_{1+d}Te_{1-x}Se_x$ is partially deintercalated by the alcoholic beverage treatment and the content of $d$ is closely correlated with the superconductivity in Fe chalcogenides.

## 4. Conclusion

We found that alcoholic beverage treatments were effective in inducing superconductivity in $Fe_{1+d}Te_{1-x}Se_x$. The most effective liquid among the tested alcoholic beverages is red wine: the $T_c^{mag}$ and the shielding volume fractions are about 10.4 K and 64.3 %. From the ICP measurements, it could be seen that part of the excess Fe was deintercalated from the interlayer sites. This behavior is similar to the case of $Fe_{1+d}Te_{1-x}S_x$. Thus the technique can be generally applied to layered Fe-chalcogenide superconductors.

**Acknowledgement**

This work was partly supported by a Grant-in-Aid for Scientific Research (KAKENHI). This research was partly supported by Strategic International Collaborative Research Program (SICORP-EU-Japan) and Advanced Low Carbon Technology Research and Development Program (JST-ALCA), Japan Science and Technology Agency.

**Figure captions**

Figure 1. The temperature dependence of magnetic susceptibility for the as-grown sample and the samples heated in various liquids.

Figure 2. The shielding volume fraction obtained from the samples heated in various liquids as a function of pH.

Figure 3. The powder X-ray diffraction patterns for the as-grown sample and samples heated in various liquids.

Figure 4. Relationship between the molar fraction of dissolved excess Fe from the sample and the shielding volume fraction for the treated samples.

Figure 1

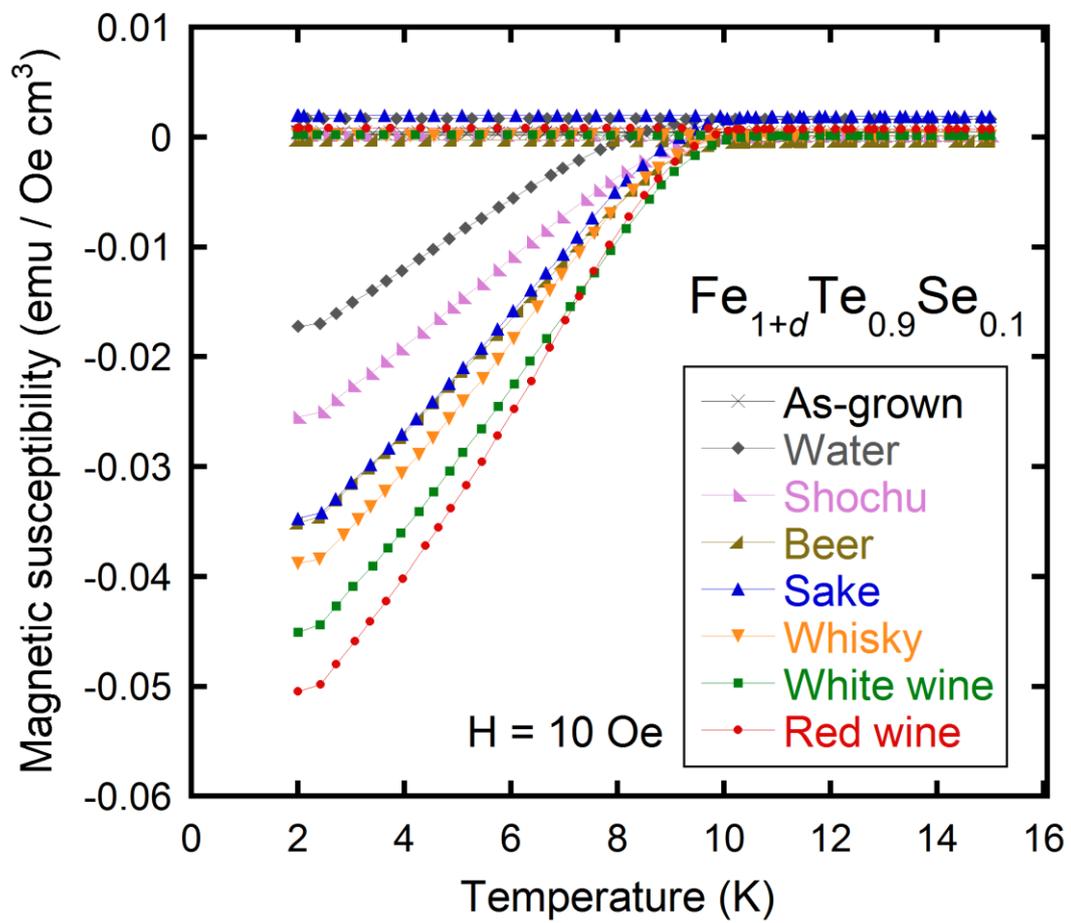

Figure 2

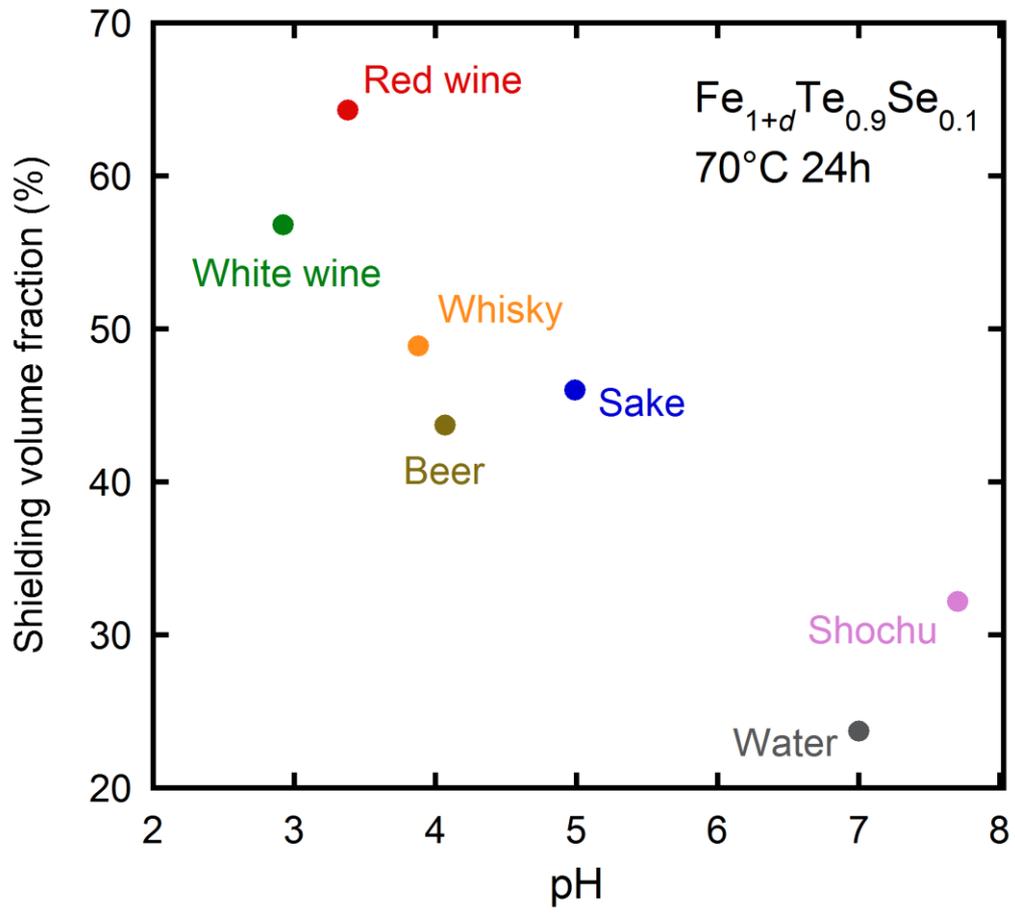

Figure 3

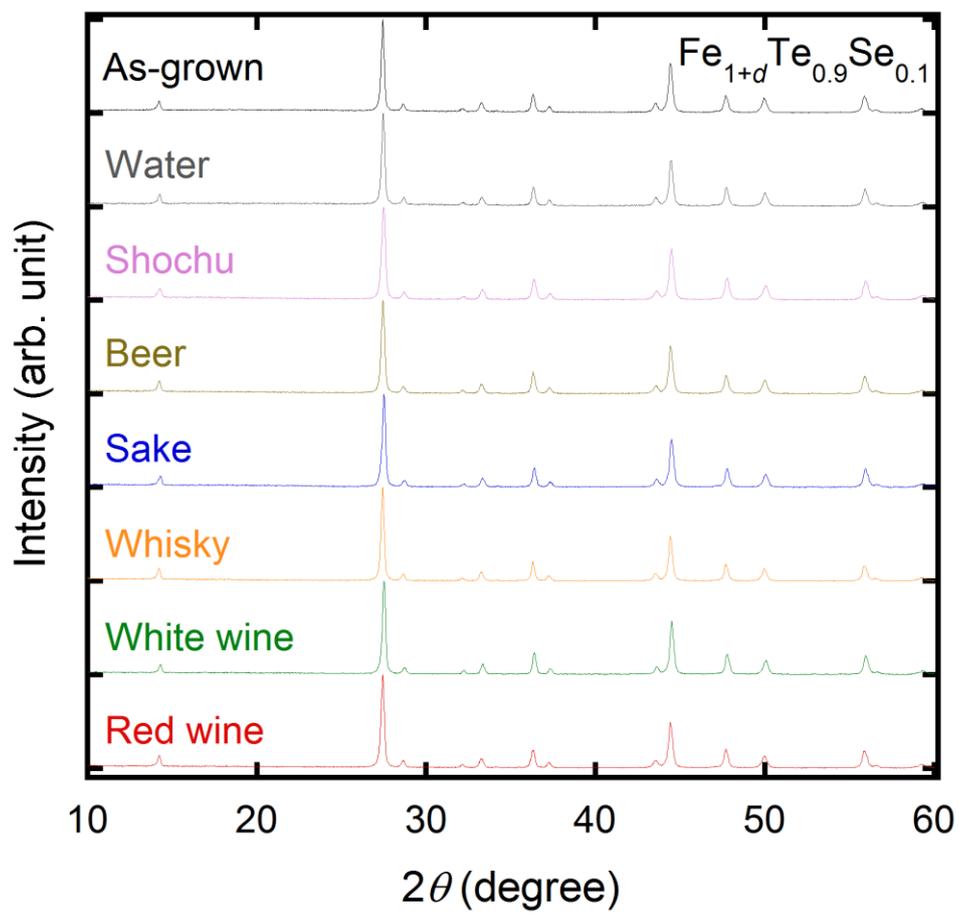

Figure 4

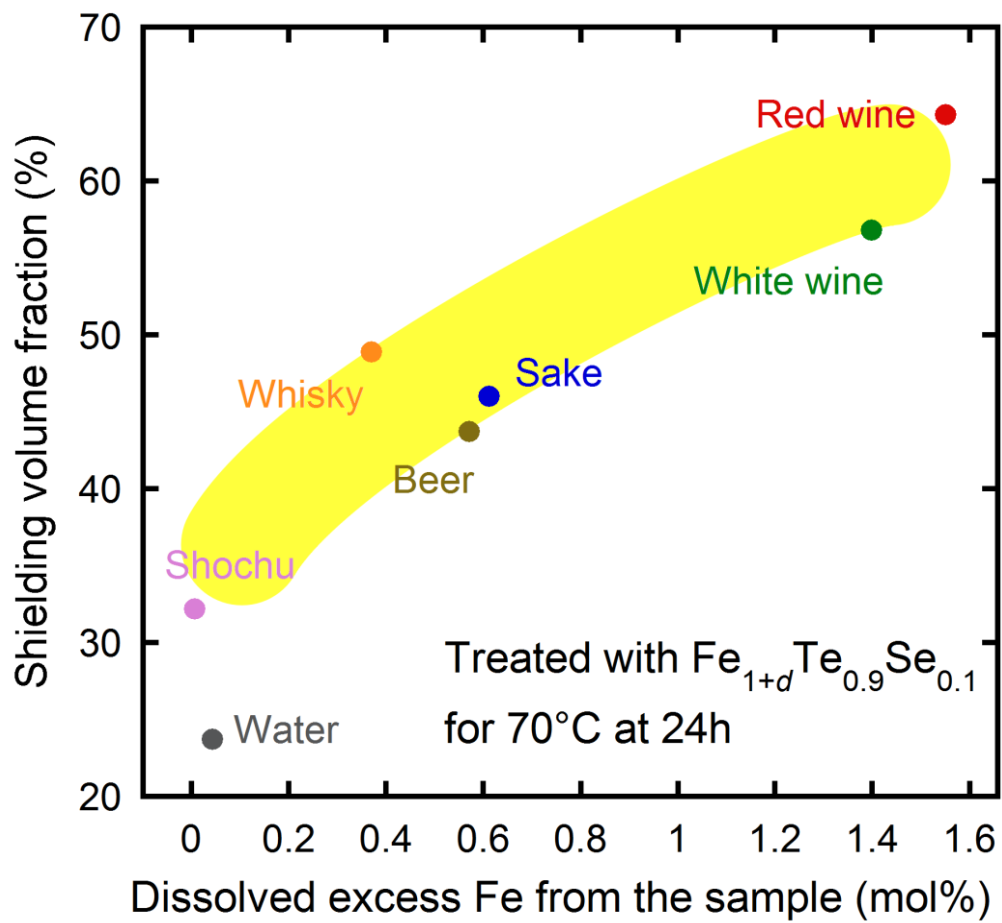